\documentclass[
aps,
twocolumn,
superscriptaddress,
nofootinbib,
floatfix,
]{revtex4-2}

\usepackage{amsmath,amssymb}
\usepackage{graphicx}
\usepackage{bm}
\usepackage{hyperref}
\hypersetup{
  colorlinks=true,
  linkcolor=blue,
  citecolor=blue,
  urlcolor=blue,
}

\usepackage{soul}
\usepackage{xcolor}
\usepackage{enumitem}
\setstcolor{red}

\begin{document}

\title{Renormalization-Group Foundation of Critical Modes in Correlation Matrices}

\author{Yanhua Zhang}
\affiliation{Department of Physics and Electronic Engineering, Yuncheng University, Yuncheng, 044000, Shanxi, China}

\author{Mingmei Xu}
\affiliation{Key Laboratory of Quark and Lepton Physics (MOE) and Institute of Particle Physics, Central China Normal University, Wuhan, 430079, Hubei, China}

\author{Jinghua Fu}
\affiliation{Key Laboratory of Quark and Lepton Physics (MOE) and Institute of Particle Physics, Central China Normal University, Wuhan, 430079, Hubei, China}

\author{Yuanfang Wu}
\email{wuyf@ccnu.edu.cn; xumm@ccnu.edu.cn; fujh@ccnu.edu.cn.}
\affiliation{Key Laboratory of Quark and Lepton Physics (MOE) and Institute of Particle Physics, Central China Normal University, Wuhan, 430079, Hubei, China}

\date{\today}

\begin{abstract}
Based on the relevant scaling operators of renormalization-group theory, we establish a theoretical framework for constructing complete families of critical-sensitive fluctuation observables and their connected correlation matrices. The corresponding eigenstates of the correlation matrices reveal a hierarchy of multiscale critical correlation patterns and the leading normalized eigenvalues and their ratios show the finite-size scaling behavior. Applying this framework to the
two-dimensional Ising model, the correlation matrices associated with both spin and energy operators are constructed. The resulting energy-sector eigenstate patterns, together with the scaling behavior of the leading eigenvalues, provide further support for the proposed framework. Finally, we discuss its potential extension to systems with unknown Hamiltonians.
\end{abstract}


\maketitle

\section{Introduction}
\label{sec:intro}

Critical phenomena are of fundamental interest across a broad range of physical systems, from condensed matter~\cite{Vojta2003,Gegenwart2008} and statistical physics to high-energy nuclear physics~\cite{Bzdak2020,Stephanov1999,Stephanov2009} and other complex systems. A defining feature of continuous phase transitions is the divergence of the correlation length in the thermodynamic limit, giving rise to long-range correlations and scale-invariant spatial fluctuations. Near a critical point, fluctuations extend over a wide range of length scales and exhibit self-similar structures, reflecting the emergence of collective behavior that is largely independent of microscopic details.

The renormalization-group (RG) theory~\cite{wilson1971renormalization,ma2001modern,nelson2025renormalization,fisher1998renormalization} provides a fundamental framework for understanding these universal properties. Through successive coarse-graining transformations, RG identifies the scaling operators that govern the long-wavelength behavior of a system, among which the relevant operators determine the dominant critical fluctuations and the associated universal scaling laws.

Experimentally identifying critical phenomena relies primarily on observables sensitive to the growth of the correlation length. Conventional approaches include susceptibilities, higher-order cumulants~\cite{Stephanov2009,Ejiri2006}, correlation functions~\cite{Bzdak2017}, and various fluctuation measures~\cite{Bialas1986,Wu1993}, whose critical behavior is characterized by universal scaling relations. These observables have been extensively employed in studies of equilibrium phase transitions and in searches for critical phenomena in complex dynamical systems.

However, many systems of physical interest, particularly those undergoing rapid evolution, may not reach thermal equilibrium. In such cases, experimentally accessible fluctuations generally contain contributions from both critical correlations and noncritical dynamical processes. Their interplay can significantly modify the expected equilibrium critical behavior~\cite{Mukherjee2015,Wu2019,Li2022}, making it difficult to isolate the underlying critical fluctuations. Developing approaches capable of identifying the characteristic spatial structures of critical correlations, without relying on thermal-equilibrium assumptions, is therefore an important challenge.

The eigen-microstate approach (EMA)~\cite{Hu2019,Sun2021}, closely related to principal component analysis (PCA)~\cite{Wang2016Discovering}, has recently provided a complementary perspective on this problem. Instead of characterizing fluctuations solely through conventional statistical observables, EMA constructs a correlation matrix from an ensemble of spatially resolved configurations and extracts its collective modes through spectral decomposition. The corresponding eigenvalues quantify the relative contributions of different modes, while their eigenvectors reveal the spatial structures of correlated fluctuations.

Previous studies of the Ising model and other critical systems have demonstrated that a small number of leading modes can develop characteristic spatial patterns associated with long-range correlations~\cite{Hu2019,Sun2021,Wang2016Discovering}. More recently, we extended EMA to relativistic heavy-ion collisions~\cite{Guo2026Eigen}, where the search for the possible critical point of quantum chromodynamics (QCD) is complicated by the short lifetime and rapid dynamical evolution of the collision system~\cite{Berdnikov2000}. By analyzing event-by-event particle fluctuations in transverse-momentum space, we demonstrated that critical fluctuations introduced into noncritical collision backgrounds can generate characteristic cluster-like eigenmode patterns and enhance the corresponding leading eigenvalues.

To connect PCA and EMA to RG theory and scaling behavior, previous studies have explored the relationship between correlation-matrix spectra and RG transformations. Bradde and Bialek~\cite{Bradde2017PCAMeetsRG} established an analogy between PCA-based spectral coarse-graining and RG transformations, while Liu et al.~\cite{Liu2022RG} developed an RG scaling description of eigen microstates and demonstrated its ability to identify critical points and critical exponents. These studies provide important insights into the relationship between spectral decomposition and critical scaling.

Despite these developments, the fundamental theoretical origin of the critical modes extracted from correlation matrices remains insufficiently understood. In particular, not all spatially resolved observables can reveal critical correlation patterns, even near a critical point. This raises the need to establish the conditions for constructing appropriate critical-sensitive observables and their connected correlation matrices, to determine the complete families of independent critical correlation modes, and to understand their hierarchical spatial structures and scaling properties. Addressing these issues requires a theoretical foundation rooted in the relevant scaling operators of RG theory.

In this work, we establish an RG-based theoretical framework for constructing critical-sensitive fluctuation observables and their connected correlation matrices. For a system with $r$ relevant scaling operators, we construct $r$ corresponding families of observables, correlation matrices, and eigenstates. RG scaling provides the basis for identifying the critical long-wavelength contributions, while singular-value decomposition (SVD) resolves the correlation spectra into a hierarchy of spatial modes, with the leading modes capturing the strongest critical fluctuations under appropriate conditions and subsequent modes revealing additional hierarchical spatial structures. We further derive the finite-size scaling behavior of the leading normalized eigenvalues and their ratios. 

Applying this framework to the two-dimensional (2D) Ising model, we construct the correlation matrices associated with both spin and energy operators. In particular, the energy-sector eigenstates reveal critical spatial correlation patterns complementary to those previously obtained in the spin sector, while the leading eigenvalues exhibit the predicted scaling behavior. Finally, we discuss the potential extension of this framework to systems with unknown Hamiltonians.

\section{RG FRAMEWORK FOR CRITICAL CORRELATION EIGENMODES}

Consider a system described by a Hamiltonian $\mathcal{H}[x]$. Suppose that, in the vicinity of a RG fixed point, the linearized RG transformation possesses $r$ relevant scaling operators $\mathcal{O}_\alpha(x)$, $\alpha=1,\ldots,r$, satisfying
\begin{equation}
 R_b\mathcal O_\alpha=b^{y_\alpha}\mathcal O_\alpha,\qquad y_\alpha>0.
\end{equation}
For each $\mathcal O_\alpha$, choose a local or quasi-local, coarse-grainable observable $A_i^{(\alpha)}$ in the same symmetry sector and with nonzero projection onto $\mathcal O_\alpha$. Define
\begin{equation}
 \delta A_i^{(\alpha)}(t)=A_i^{(\alpha)}(t)-\langle A_i^{(\alpha)}\rangle,
\end{equation}
where $i=1,\ldots,N_\alpha$ labels spatial degrees of freedom and $t=1,\ldots,M$ labels sampled configurations. Here $N_\alpha$ is the number of spatial observables in channel $\alpha$. $\langle\dots \rangle$ is the average over the sample. The connected spatial correlation is
\begin{equation}
 G_{ij}^{(\alpha)}=\langle\delta A_i^{(\alpha)}\delta A_j^{(\alpha)}\rangle.
\end{equation}
We assume that the sampled configurations are representative of a well-defined statistical ensemble, so that the empirical connected correlations converge to their ensemble counterparts in the large-sample limit. This construction does not require thermal equilibrium, although the interpretation of critical scaling properties relies on the corresponding RG
scaling assumptions.

Define the normalized $N_\alpha\times M$ data matrix
\begin{equation}
X^{(\alpha)}=\frac{1}{\sqrt{\mathcal N_\alpha}}\begin{pmatrix}
\delta A_1^{(\alpha)}(1)&\delta A_1^{(\alpha)}(2)&\cdots&\delta A_1^{(\alpha)}(M)\\
\delta A_2^{(\alpha)}(1)&\delta A_2^{(\alpha)}(2)&\cdots&\delta A_2^{(\alpha)}(M)\\
\vdots&\vdots&\ddots&\vdots\\
\delta A_{N_{\alpha}}^{(\alpha)}(1)&\delta A_{N_{\alpha}}^{(\alpha)}(2)&\cdots&\delta A_{N_{\alpha}}^{(\alpha)}(M)
\end{pmatrix},
\end{equation}
with
\begin{equation}
 X_{it}^{(\alpha)}=\frac{\delta A_i^{(\alpha)}(t)}{\sqrt{\mathcal N_\alpha}},\qquad
 \mathcal N_\alpha=\sum_{i=1}^{N_\alpha}\sum_{t=1}^{M}[\delta A_i^{(\alpha)}(t)]^2.
\end{equation}
Where $\mathcal N_\alpha$ is the global normalization factor. Define the sampled-configuration and spatial-variable correlation matrices, respectively, by
\begin{align}
 C^{(\alpha)}&=X^{(\alpha)T}X^{(\alpha)}\quad(M\times M),\\
 K^{(\alpha)}&=X^{(\alpha)}X^{(\alpha)T}\quad(N_\alpha\times N_\alpha).
\end{align}
With sufficiently complete representative sampling, $K_{ij}^{(\alpha)}\propto G_{ij}^{(\alpha)}$.


Let
\begin{equation}
X^{(\alpha)}=U^{(\alpha)}\Sigma^{(\alpha)}V^{(\alpha)T}
\end{equation}
be its singular-value decomposition, with nonzero singular values $\sigma_1^{(\alpha)}\geq\sigma_2^{(\alpha)}\geq\cdots\geq0$ as the diagonal elements of matrix $\Sigma^{(\alpha)}$ with rank $q_\alpha\leq\min(N_\alpha,M)$. Then
\begin{equation}
 K^{(\alpha)}U_I^{(\alpha)}=(\sigma_I^{(\alpha)})^2U_I^{(\alpha)},\ \ 
 C^{(\alpha)}V_I^{(\alpha)}=(\sigma_I^{(\alpha)})^2V_I^{(\alpha)}.
\end{equation}
$U^{(\alpha)}$ and $V^{(\alpha)}$ are $N_\alpha\times N_\alpha$, $M\times M$ matrices, composed by eigenvectors $U_I^{(\alpha)}$ and $V_I^{(\alpha)}$ respectively. The two correlation matrices share identical nonzero eigenvalues, while their corresponding eigenvectors are related through the singular-value decomposition. Specifically, 
\begin{equation}
X^{(\alpha)}V_I^{(\alpha)}=\sigma_I^{(\alpha)}U_I^{(\alpha)},\ 
\mathrm{and}\ \ X^{(\alpha)T}U_I^{(\alpha)}=\sigma_I^{(\alpha)}V_I^{(\alpha)}.
\end{equation}
Equations (9) and (10) establish the spectral equivalence between the two correlation matrices: they possess identical nonzero eigenvalues, while their corresponding eigenstates are related by the SVD mapping. Thus, the spatial eigenstates of $K^{(\alpha)}$ can be reconstructed from those of $C^{(\alpha)}$, and vice versa. In particular, the largest eigenvalues of the two matrices are identical. Since $\mathrm{tr}\,K^{(\alpha)}=\mathrm{tr}\,C^{(\alpha)}=1$, the normalized eigenvalue weights are defined as
\begin{equation}
 w_I^{(\alpha)}=(\sigma_I^{(\alpha)})^2,\qquad
 \sum_{I=1}^{q_\alpha}w_I^{(\alpha)}=1.
\end{equation}
Within this RG-based framework, we can obtain following three principal results.

\vspace{0.8em}
\noindent\textbf{(I) Critical-Sensitive Observables and Correlation-Matrix Families.}
\vspace{0.6em}

The scaling-operator expansion of the chosen centered observable contains
\begin{equation}
    \delta A^{(\alpha)}
=a_\alpha \mathcal{O}_\alpha
+\sum_{\beta\ne\alpha}a_\beta \mathcal{O}_\beta
+\sum_\gamma b_\gamma \mathcal{O}_\gamma^{\mathrm{irr}},
\quad a_\alpha\ne0.
\end{equation}
The expansion includes scaling contributions allowed by symmetry, with coefficients determined by the choice of observable. The nonzero projection onto the relevant scaling operator $\mathcal{O}_\alpha$ ensures that the corresponding connected correlation function contains its critical long-distance scaling contribution. Together with locality, coarse-grainability, appropriate symmetry properties, and representative statistical sampling, this provides the theoretical basis for constructing critical-sensitive observables and their connected correlation matrices.

For a system with $r$ independent RG relevant scaling operators, a complete set of $r$ independent observable families can be constructed by choosing observables whose projections span the relevant scaling-operator space. Each family defines a corresponding connected correlation matrix and its associated eigenstate spectrum. The resulting $r$ families provide a systematic representation of the critical correlation modes associated with the relevant scaling operators.

\vspace{0.8em}
\noindent\textbf{(II) Multiscale Hierarchy of Critical Correlation Modes.}
\vspace{0.6em}

Near criticality, the growth of the correlation length generates enhanced spatial correlations over a broad range of length scales. RG coarse-graining identifies the relevant scaling contributions governing these long-wavelength fluctuations, while progressively suppressing the contributions of irrelevant operators at large scales.

The SVD of the centered data matrix resolves the resulting connected correlations into orthogonal spatial eigenmodes, ordered by their eigenvalue weights. For critical correlations with long-wavelength spectral enhancement, larger spatial scales carry stronger correlation weights. Consequently, the leading eigenstates are dominated by long-wavelength collective fluctuations, while subsequent eigenstates progressively resolve weaker correlations with finer characteristic spatial structures.

This establishes a multiscale hierarchy of critical correlation modes, in which the decreasing eigenvalue weights are associated with a progression from stronger, large-scale collective correlations to weaker, shorter-wavelength structures. The resulting eigenstate patterns exhibit extended coherent regions in the leading modes and increasingly refined patch-like or cluster-like structures in higher-order modes, reflecting the hierarchical organization of critical fluctuations across spatial scales.

\vspace{0.8em}
\noindent\textbf{(III) Finite-Size Scaling of Leading Eigenvalue Weights.}
\vspace{0.6em}

For the normalized connected correlation matrix associated with an RG scaling operator $\mathcal{O}_\alpha$, the largest eigenvalue $w_1^{(\alpha)}$ measures the relative strength of the dominant correlation mode within the corresponding observable sector.

Near criticality, the connected correlation function of the corresponding scaling operator behaves as
\begin{equation}
    G^{(\alpha)}(l)\sim l^{-2x_\alpha},
\end{equation}
where $l$ denotes the spatial separation and $x_\alpha=d-y_\alpha$ is the scaling dimension of $\mathcal{O}_\alpha$.

For a translationally invariant system whose leading eigenmode is governed by the long-wavelength critical contribution, the normalized leading eigenvalue weight is proportional to the spatially integrated connected correlation function divided by the total local fluctuation strength. Assuming a finite and nonzero local fluctuation variance in the thermodynamic limit, we obtain
\begin{equation}
    w_1^{(\alpha)}(L)
\sim \frac{1}{L^d}
\int_a^L l^{d-1}G^{(\alpha)}(l)\,\mathrm{d}l,
\end{equation}
where $a$ denotes a microscopic length scale.

Consequently,
\begin{equation}
w_1^{(\alpha)}(L)\sim
\begin{cases}
L^{-2x_\alpha}, & 2x_\alpha<d,\\[2mm]
L^{-d}\ln(L), & 2x_\alpha=d,\\[2mm]
L^{-d}, & 2x_\alpha>d.
\end{cases}
\end{equation}
Thus, the finite-size scaling of the leading eigenvalue weight is governed by the scaling dimension of the corresponding RG operator, with a logarithmic correction at the borderline condition $2x_\alpha=d$~\cite{kenna2006selfconsistent,salas2001exact}.

For two scaling operators satisfying $2x_\alpha<d$ and $2x_\beta<d$, the ratio of their leading eigenvalue weights obeys
\begin{equation}
\frac{w_1^{(\alpha)}(L)}
{w_1^{(\beta)}(L)}
\sim L^{2(y_\alpha-y_\beta)}. 
\end{equation}
When either scaling operator satisfies $2x=d$, the corresponding logarithmic correction enters the ratio.

At a given system size and thermodynamic state, this ratio characterizes the relative strengths of the dominant correlation modes associated with different RG scaling operators, measured by their normalized spectral weights. Its finite-size scaling further reflects the difference between their scaling dimensions, providing a quantitative measure of the relative critical-mode strengths across the corresponding observable sectors.

\section{Applications to 2D-Ising model}

The established RG-based framework constructs critical-sensitive observables and their connected correlation matrices, reveals the multiscale hierarchy of critical correlation modes, and predicts the finite-size scaling of their leading eigenvalue weights. The 2D Ising model provides a suitable setting for examining these results, since its relevant RG scaling operators and scaling dimensions are well established~\cite{cardy1996scaling,difrancesco1997conformal}. In particular, the spin and energy operators allow us to investigate two independent families of critical correlation modes and their corresponding spatial hierarchies and scaling properties.

\begin{figure*}[t]
\centering
\includegraphics[width=\linewidth]{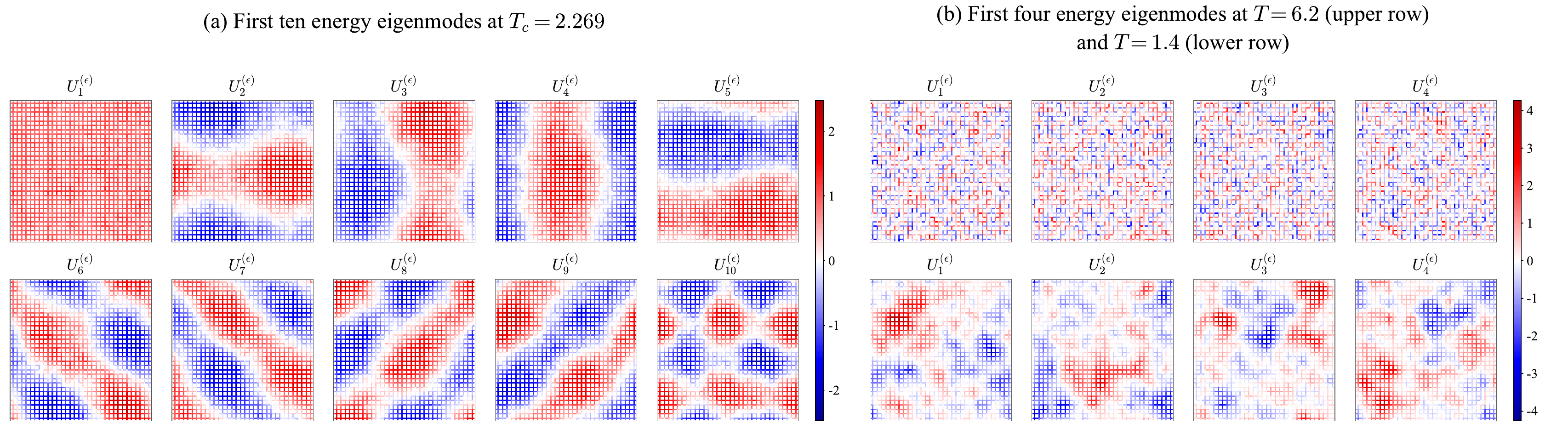}
\caption{Spatial patterns of the energy-sector eigenmodes of the 2D Ising model for $L=32$ and $M=20{,}000$. (a) The first ten eigenmodes at $T_c=2.269$. (b) The first four eigenmodes at $T=6.2$ (upper row) and $T=1.4$ (lower row).}
\end{figure*}

For a square lattice of linear size $L$ with periodic boundary conditions, the Hamiltonian of the 2D Ising model is
\begin{equation}
\mathcal H=-J\sum_{\langle i,j\rangle}s_i s_j-h\sum_i s_i,\qquad s_i=\pm1,
\end{equation}
where $\langle i,j\rangle$ denotes nearest-neighbor sites. At the zero-field Ising critical fixed point, the two relevant bulk scaling operators are the spin operator $\mathcal O_s$ and the energy-density operator $\mathcal O_\epsilon$, with scaling dimensions $x_s=1/8$ and $x_\epsilon=1$, respectively~\cite{cardy1996scaling,difrancesco1997conformal}. Their RG scaling exponents are $y_s=2-x_s=15/8$ and $y_\epsilon=2-x_\epsilon=1$. According to the first principal result of the
proposed framework, these two operators motivate two separate sets of centered fluctuation observables and their corresponding connected correlation matrices.

The spin and energy operators belong to the $Z_2$-odd and $Z_2$-even symmetry sectors, respectively, allowing their critical contributions to be distinguished by symmetry. The local spin
and nearest-neighbor bond-energy observables have nonvanishing overlap with the corresponding RG scaling operators, while centering removes their constant contributions. Moreover, the translational invariance of the equilibrium system with periodic boundary conditions, together with the finite and nonzero local fluctuation strengths, provides the conditions for examining the predicted finite-size scaling of their normalized leading eigenvalue weights.

For the spin scaling operator, we choose the local spin variables $A_i^{(s)}=s_i$ at the $N_s=L^2$ lattice sites. The centered fluctuation observables and their connected correlation matrix are
\begin{equation}
    \delta A_i^{(s)}=s_i-\langle s_i\rangle,
\end{equation}
and 
\begin{equation}
    G_{ij}^{(s)}=\langle \delta A_i^{(s)}\delta A_j^{(s)}\rangle
=\langle s_i s_j\rangle-\langle s_i\rangle\langle s_j\rangle,
\quad i,j=1,\ldots,L^2.
\end{equation}
The normalized spatial correlation matrix $K^{(s)}$, constructed from these spin fluctuations, is the matrix underlying the spatial eigenmode analysis of the two-dimensional Ising model. The previously reported critical spin-sector eigenstate patterns can therefore be interpreted within the present RG framework as spatial representations of fluctuations associated with the spin scaling operator. However, these studies did not construct the corresponding connected correlation matrix and eigenstate patterns for the energy scaling operator. 

To address the energy sector, we label each nearest-neighbor bond by $a=(i,\mu)$, where $\mu=x,y$ denotes the horizontal or vertical direction, and define $e_{i,\mu}=s_i s_{i+\hat\mu}$. The centered energy fluctuation observables are
\begin{equation}
\delta A_{i,\mu}^{(\epsilon)}=e_{i,\mu}-\langle e_{i,\mu}\rangle.
\end{equation}
With periodic boundary conditions, there are $N_\epsilon=2L^2$ nearest-neighbor bonds, compared with $N_s=L^2$ spin observables. The corresponding unnormalized spatial connected correlation matrix $G^{(\epsilon)}$ therefore has dimensions $2L^2\times 2L^2$, with elements
\begin{align}
G_{(i,\mu),(j,\nu)}^{(\epsilon)}
&=\langle \delta A_{i,\mu}^{(\epsilon)} \delta A_{j,\nu}^{(\epsilon)}\rangle\notag\\
&=\langle e_{i,\mu}e_{j,\nu}\rangle
-\langle e_{i,\mu}\rangle\langle e_{j,\nu}\rangle.
\end{align}
For the numerical analysis, we construct the normalized $N_\alpha\times N_\alpha$ matrix $K^{(\epsilon)}$ from the centered bond-energy fluctuations, and obtain the corresponding spatial eigenstates $U_I^{(\epsilon)}$. 


Each energy-sector eigenstate $U_I^{(\epsilon)}$ contains $2L^2$ components corresponding to the horizontal and vertical nearest-neighbor bonds. In Fig.~1, these components are displayed at their corresponding bond positions, with colors representing the eigenvector amplitudes. A common color scale is used for the eigenstates at each temperature, allowing their spatial structures to be compared directly.

Fig.~1(a) presents the first ten energy-sector eigenstate patterns at $T_{\rm c}=2.269$, obtained with $L=32$ and $M=20{,}000$ equilibrium configurations generated by the Wolff algorithm. The leading eigenstate exhibits a nearly uniform spatial structure, reflecting a collective long-wavelength correlation mode. The subsequent eigenstates develop extended regions of opposite signs, which progressively evolve into more numerous and finer patches with increasing mode index. This systematic progression reveals a clear hierarchy of spatial correlation structures, from the strongest large-scale collective mode to weaker modes with finer characteristic spatial variations. The observed hierarchy closely resembles that of the critical spin-sector eigenstates and provides numerical support for the second principal result: the multiscale critical correlations associated with an RG relevant operator are resolved into a hierarchy of spatial eigenmodes ordered by their correlation weights.

Fig.~1(b) shows the first four energy-sector eigenstates at $T=6.2$ (upper row) and $T=1.4$ (lower row). At both temperatures, the spatial patterns are less coherent than those at criticality and exhibit more irregular, fragmented structures. The systematic development of extended correlation regions into progressively finer spatial patterns is most pronounced at $T_{\rm c}$, reflecting the enhancement of long-wavelength correlations near the critical point. Together, these results demonstrate that the energy scaling operator generates its own family of critical spatial correlation modes, with a hierarchical organization consistent with the RG-based theoretical framework.

\begin{figure}[t]
\centering
\includegraphics[width=\linewidth]{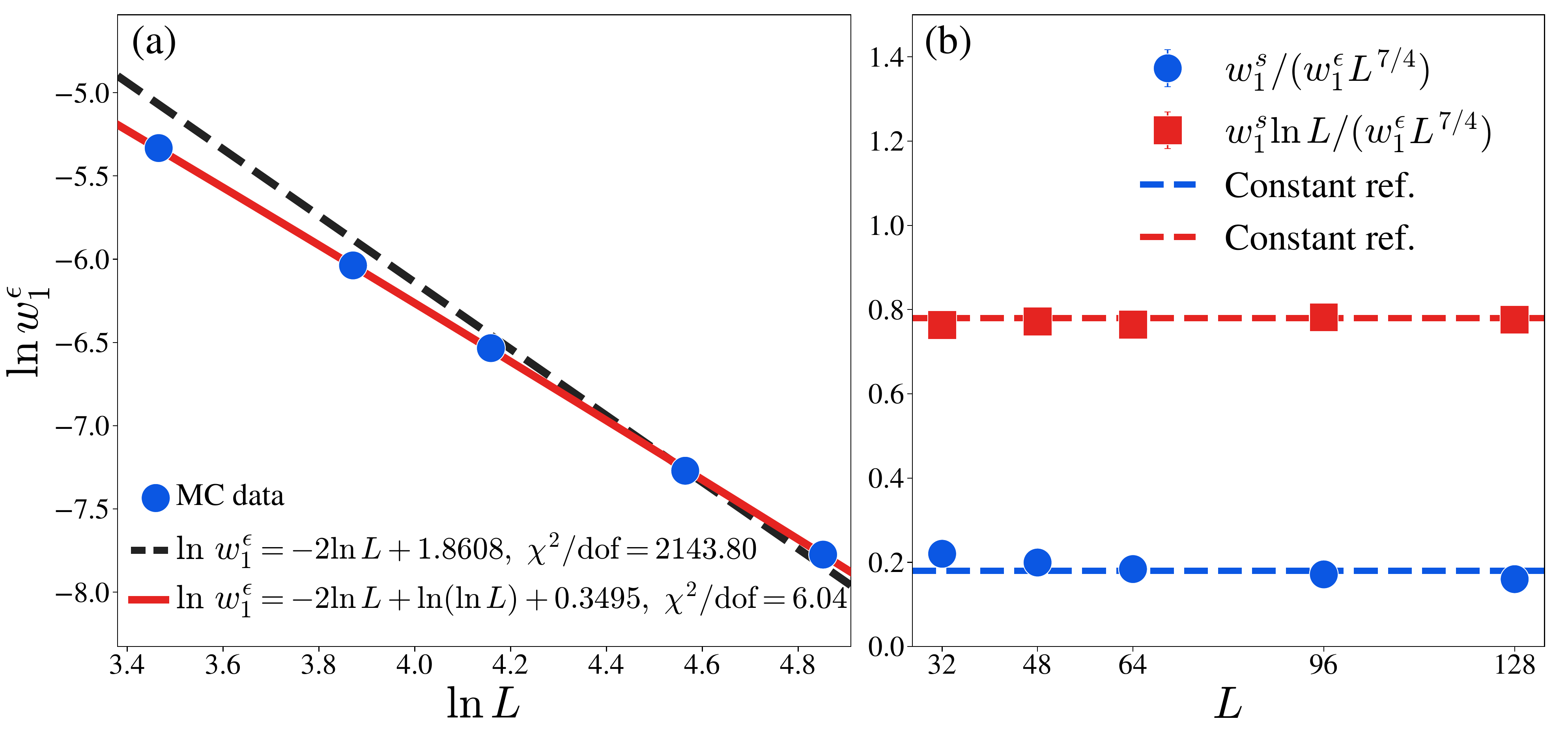}
\caption{Finite-size scaling of the leading eigenvalue weights at $T_{\rm c}$ for $L=32,48,64,96$, and $128$. (a) Log-log plot of the leading energy-sector eigenvalue weight $w_1^{(\epsilon)}$. (b) Ratios $w_1^{(s)}/(w_1^{(\epsilon)}L^{7/4})$ (blue circles) and $w_1^{(s)}\ln L/(w_1^{(\epsilon)}L^{7/4})$    (red squares). The dashed lines indicate constant reference values. }
\end{figure}

We next examine the finite-size scaling of the leading eigenvalue weights in the spin and energy sectors at $T_{\rm c}$, using five lattice sizes, $L=32,48,64,96$, and 128. For the spin sector, a power-law fit yields $w_1^{(s)}\propto L^{-0.243}$, in excellent agreement with the RG prediction $w_1^{(s)}\sim L^{-1/4}$ and the previously reported exponent of 0.246(6)~\cite{Hu2019}.

For the energy sector, the scaling dimension $x_\epsilon=1$ satisfies the borderline condition $2x_\epsilon=d=2$, leading to the RG prediction $w_1^{(\epsilon)}\sim L^{-2}\ln L$. Fig.~2(a) presents the Monte Carlo results for $w_1^{(\epsilon)}$, together with fits using the theoretical scaling forms with and without the logarithmic correction. Statistical uncertainties are included for all data points but are smaller than the plotted symbols. To obtain stable estimates at the two largest system sizes, $L=96$ and 128, the number of sampled configurations was increased to $M=60{,}000$.

The pure power-law form, $\ln w_1^{(\epsilon)}=-2\ln L+1.8608$, gives $\chi^2/\mathrm{dof}=2143.80$, exhibiting substantial deviations from the numerical results. In contrast, the logarithmically corrected form, $\ln w_1^{(\epsilon)}=-2\ln L+\ln(\ln L)+0.3495$, provides a much better description, with $\chi^2/\mathrm{dof}=6.04$. The remaining deviation of the reduced $\chi^2/\mathrm{dof}$ from unity is sensitive to the very small statistical uncertainties, particularly at the two largest lattice sizes. These results provide strong numerical support for the logarithmic correction predicted by RG theory for the energy scaling operator. 

Fig.~2(b) presents the ratios $w_1^{(s)}/(w_1^{(\epsilon)}L^{7/4})$ and $w_1^{(s)}\ln L/(w_1^{\epsilon)}L^{7/4})$, shown by the blue circles and red squares, respectively. The blue points exhibit a slight decreasing trend with increasing system size, whereas the logarithmically corrected red points remain approximately constant over the investigated range. This provides further support for the predicted relative scaling $w_1^{(s)}/w_1^{(\epsilon)}\sim L^{7/4}/\ln L$.


The same strategy suggests a route to more complicated systems whose microscopic Hamiltonians and relevant RG operators are not known in advance, and whose thermal equilibrium may not be established. One may construct physically motivated local or coarse-grainable fluctuation observables, form their connected correlation matrices, and examine the spatial organization and scaling behavior of the leading eigenvalues. 

Importantly, the construction of these matrices does not require thermal equilibrium, although nonequilibrium effects must be considered when interpreting their critical behavior. Such an analysis may reveal candidate long-wavelength collective modes and identify observables sensitive to critical correlations. Conversely, the resulting eigenstate patterns and scaling properties may provide constraints on the underlying effective Hamiltonian and relevant RG operators. This approach may be particularly useful in relativistic heavy-ion collisions, where the rapidly evolving system poses challenges for identifying critical fluctuations associated with the possible QCD critical point.

\section{Summary}

In summary, we have established a RG framework for constructing critical-sensitive fluctuation observables and their connected correlation matrices. Starting from the relevant RG scaling operators, we formulate the conditions under which the selected observables capture critical long-range correlations. For a system with $r$ independent relevant scaling operators, $r$ corresponding families of observables, correlation matrices, and eigenstates can be constructed. RG scaling provides the physical basis for the enhancement of long-wavelength critical correlations, while singular-value decomposition resolves these correlations into a hierarchy of spatial eigenmodes, with stronger large-scale collective modes followed by weaker modes exhibiting finer spatial structures. We further derive the finite-size scaling behavior of the leading normalized eigenvalue weights and their ratios, connecting the spectral properties of correlation matrices with the scaling dimensions of the underlying RG operators.

We apply this framework to the 2D Ising model by constructing connected correlation matrices associated with its two relevant scaling operators, spin and energy. In addition to the previously studied spin-sector patterns, we identify a distinct family of energy-sector eigenstates that exhibit a characteristic multiscale hierarchy of critical spatial correlations. The leading eigenvalue weights reveal different strengths and scaling behaviors in the two sectors. The energy-sector results provide strong numerical support for the RG-predicted logarithmic correction. The relative scaling of the leading eigenvalue weights further confirms the distinct critical scaling properties of the two RG operators. Together, these results support the proposed theoretical framework for identifying and characterizing independent families of critical correlation modes.

The framework also suggests a possible extension to systems with unknown microscopic Hamiltonians, where suitably constructed fluctuation observables may reveal critical correlation modes and provide constraints on the underlying effective description. Such an approach should be particularly useful in relativistic heavy-ion collisions for investigating critical fluctuations associated with the possible QCD critical point.


\vspace{0.5cm}
\begin{acknowledgments}
The authors thank Prof. Dr. Xiangdong Ji for helpful discussions, comments, and encouragement. The first author thanks Dr. Ranran Guo for helpful guiding in numerical calculations of the paper. This research was funded by the National Key Research and Development Program of China, Grant No. 2024YFA1610700, Grant No. 2022YFA1604900, the National Natural Science Foundation of China, Grant No. 12275102, and PhD research startup foundation of Yuncheng University (YQ-2020023); Doctor returned to Shanxi Province Project (QZX-2021006).
\end{acknowledgments}

% ...

\bibliography{ref}   

\end{document}